\documentclass[a4paper]{jpconf}
\usepackage{epsfig}

\begin{document}
\title{Search for the critical point of strongly interacting matter at the CERN SPS NA61/SHINE experiment\footnote{Talk presented at the International Workshop: Discovery Physics at the LHC -- Kruger 2014, RPA, December~1-6,~2014}}

\author{Ludwik Turko\\ for the NA61/SHINE Collaboration}
\address{Institute of Theoretical Physics, University of Wroclaw, pl. M.~Borna 9, 50-205 Wroclaw, Poland}
\ead{ludwik.turko@ift.uni.wroc.pl}


\begin{abstract}
The NA61/SHINE experiment performs a detailed study of the onset of deconfinement and search for critical point of hadronic matter by colliding nuclei of different sizes at various beam momenta from 13A to 158A GeV/c. Experimental setup and results on the theoretically expected signatures are discussed.
\end{abstract}

\section {Introduction}
The NA61/SHINE, understood as The \textbf{S}uper Proton Synchrotron (SPS) \textbf{H}eavy \textbf{I}on and
\textbf{N}eutrino \textbf{E}xperiment, is the continuation and the extension of the NA49 experiment \cite{Antoniou:2006_034,  Abgrall:2008_018}. It uses the same experimental fix target set up as NA49  but the physics programme has been extended to measure the production of charged pions and kaons out of a thin carbon target and a replica of the T2K target what is necessary to test accelerator neutrino beams \cite{Abgrall:2011ae}. Data taking began in 2007.   These  $p-C$ data allow also for better understanding of nuclear cascades in the cosmic-air showers - necessary in the in the Pierre Auger and KASCADE experiments \cite{Auger_2004, Kascade_2003}.  The main aim of the NA61/SHINE, however, is a study of the onset of deconfinement and search for critical point of hadronic matter. The NA49 experiment studied  hadron production in Pb+Pb interactions while the NA61/SHINE collect data at  varying collision energy (13A-158A GeV) and size of the colliding systems. This is equivalent to the two dimensional scan of the hadronic phase diagram in the $T, \mu_B$ plane. The ion collisions programme was initiated in 2009 with the $p-p$ collisions used later on as reference data.

\begin{figure}[!]
\includegraphics[width=0.35\textwidth]{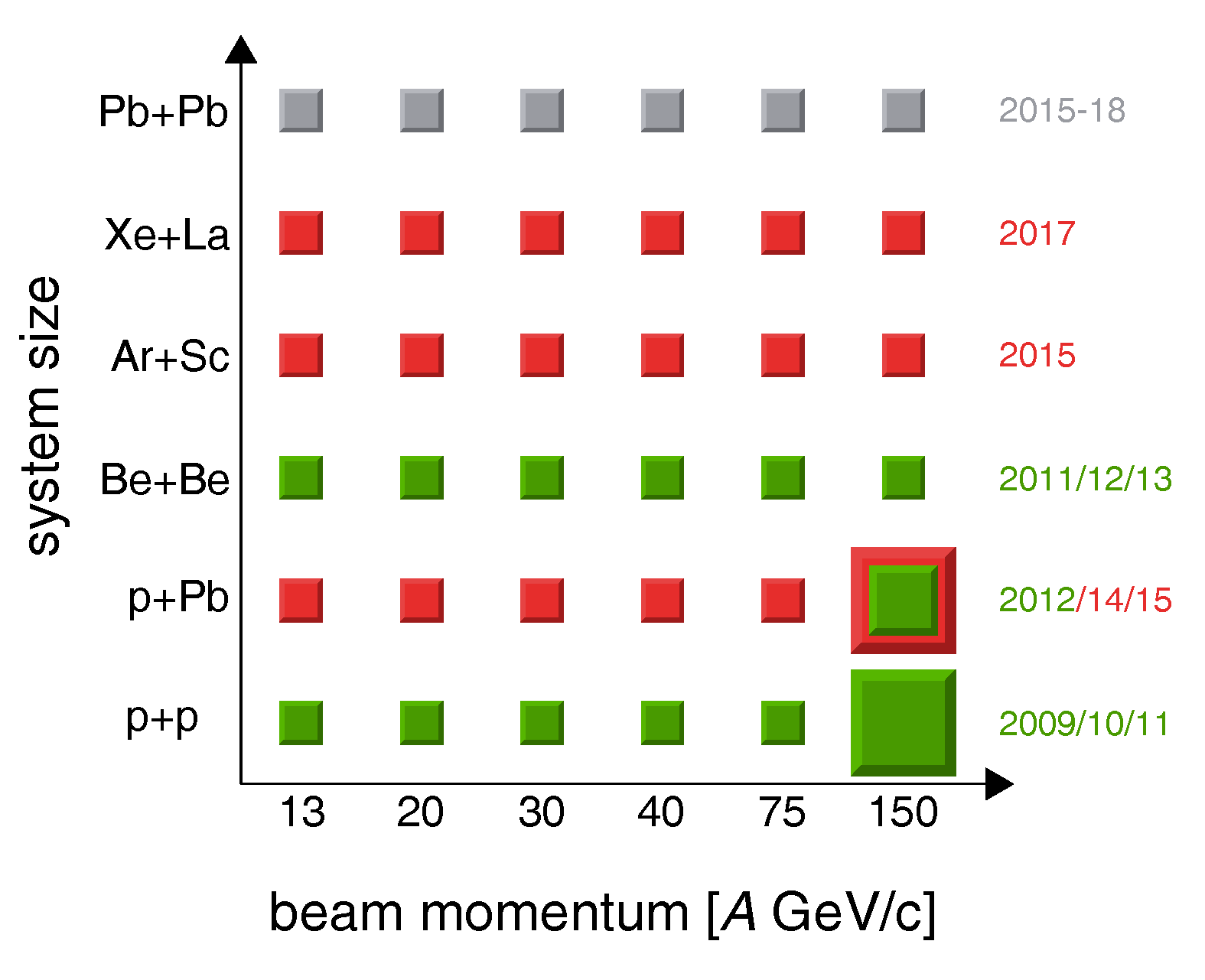}
\includegraphics[width=0.65\textwidth]{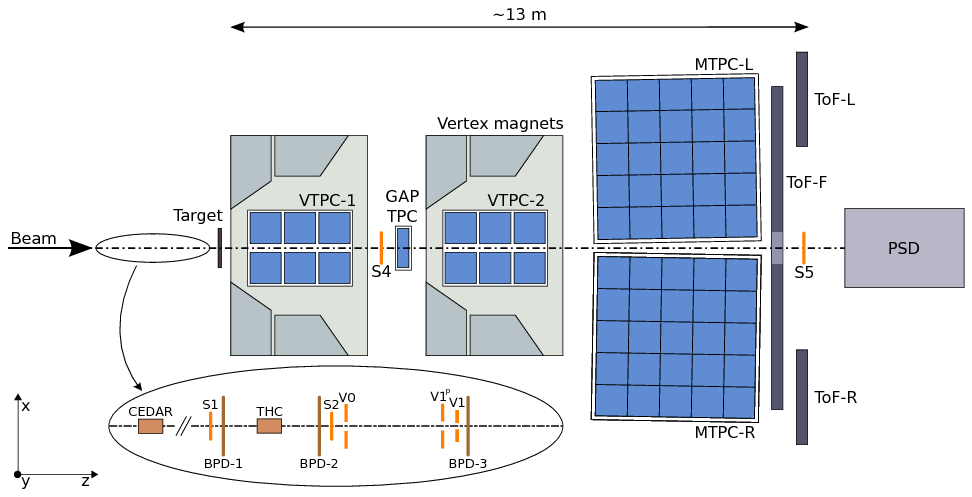}
\caption{\label{Fig.1}
\textbf{Left:} For the programme on strong interactions NA61 / SHINE scans in the system size and beam momentum.
In the plot the recorded data are indicated in green, the approved future data taking in red, whereas the proposed
extension for the period $2015–2018$ in gray.\\ \textbf{Right:} The NA61 / SHINE detector consists of a large acceptance hadron spectrometer followed by a set of six Time Projection Chambers
as well as Time-of-Flight detectors. The high resolution forward calorimeter, the Projectile Spectator Detector,
measures energy flow around the beam direction. For hadron-nucleus interactions, the collision volume is determined by counting low momentum particles emitted from the nuclear target with the LMPD detector (a small TPC) surrounding the target. An array of beam detectors identifies beam particles, secondary hadrons and nuclei as well as primary nuclei, and measures precisely their trajectories.}
\end{figure}

The programme allows to measure also fluctuations of various physical quantities which are sensitive to the phase transition. These fluctuations create  experimental signature for the critical point. An analysis of  fluctuations of various observables, particularly in a range of energies around 8 GeV at the center of mass in interactions of light nuclei per colliding nucleon pair (it corresponds to the beam energy of 30 GeV in the frame of a stationary target), is the main goal of the NA61/SHINE experiment. This is the kinematical region where NA49 data indicate that the onset of deconfinement in central Pb+Pb collisions. It is mainly based on the observation of structures in the energy dependence of
hadron production in central Pb+Pb collisions which are not observed in elementary interactions \cite{Alt:2008,Gazdzicki:2008}

The collaboration consists of about 150 physicists from 15 countries and 30 institutions. It is the second largest non-LHC experiment at CERN.
\begin{figure}[!h]
\includegraphics[width=0.4\textwidth]{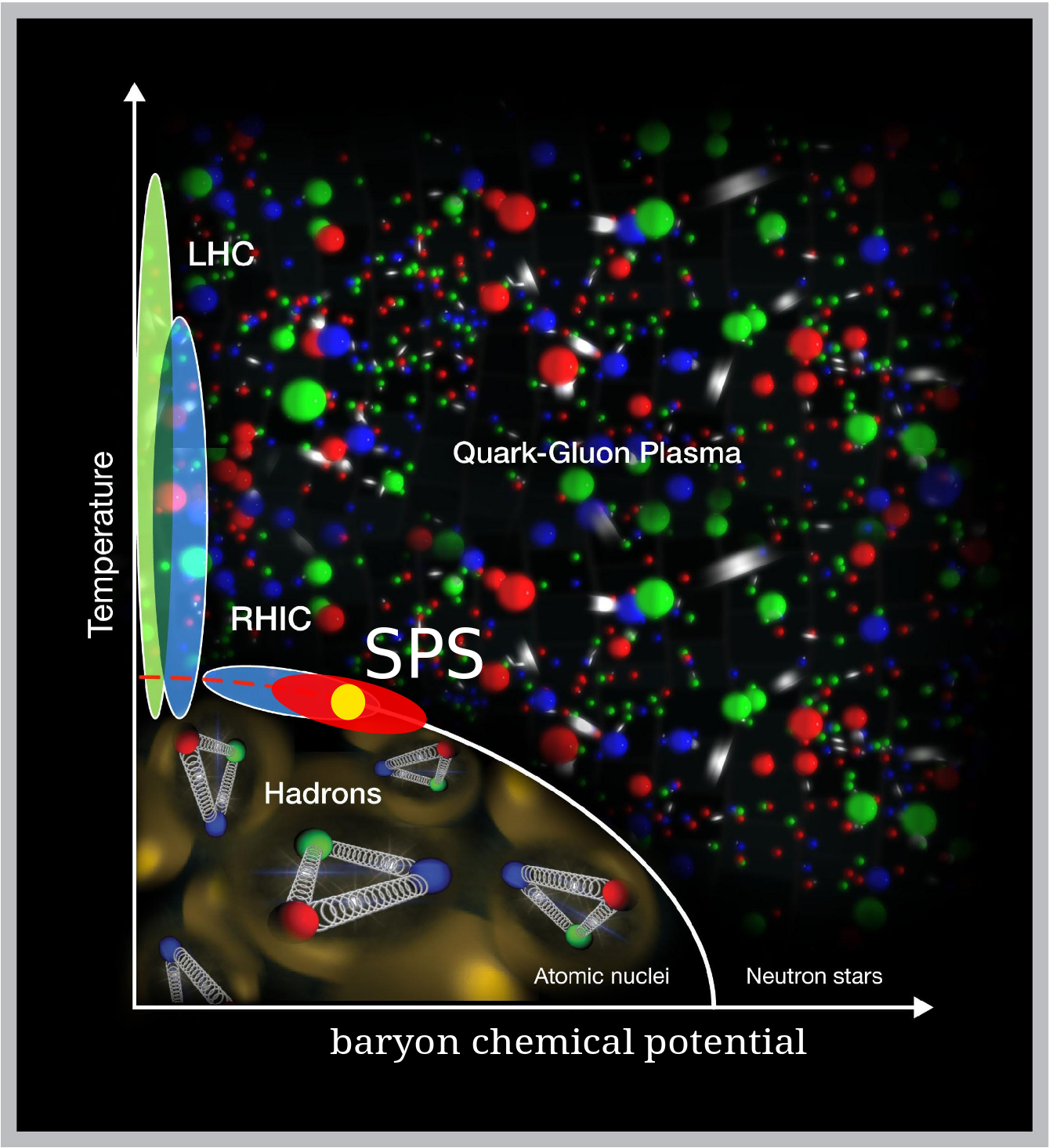}
\includegraphics[width=0.5\textwidth]{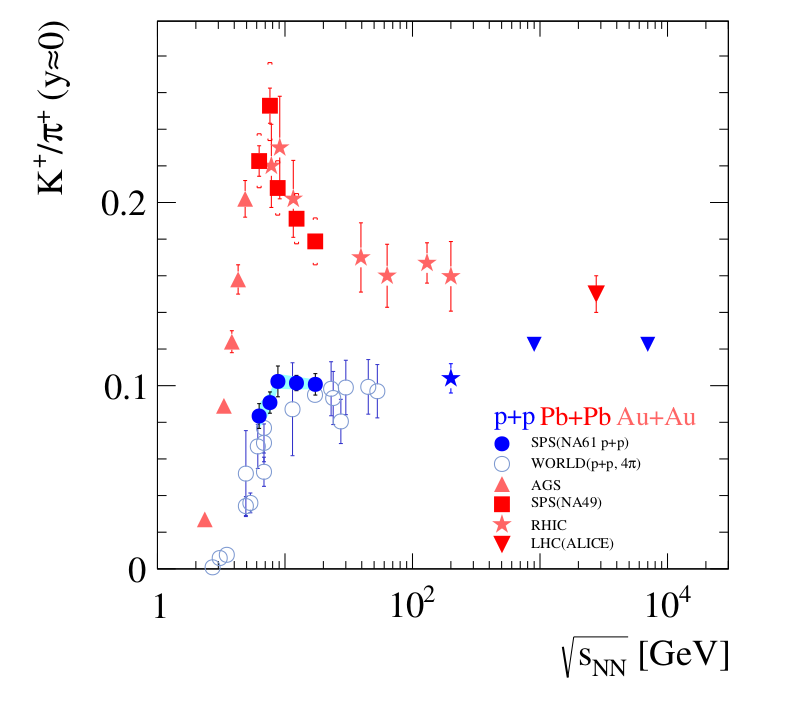}
\caption{\label{Fig.2}
 \textbf{Left:} Phase diagram of strongly interacting matter in the $( T , \mu_B)$ plane\\
 \textbf{Right:} Horn - a strong maximum of the ratio of $K^+/pi^+$ multiplicities. A reduced shadow of the horn structure is visible in $p+p$ reactions.}
\end{figure}

\section{Some new  NA61/SHINE results}
\subsubsection*{}
NA61 measurements established energy dependence of the inelastic cross section ${}^7Be + {}^9Be$

\begin{figure}[!h]
\includegraphics[width=0.5\textwidth]{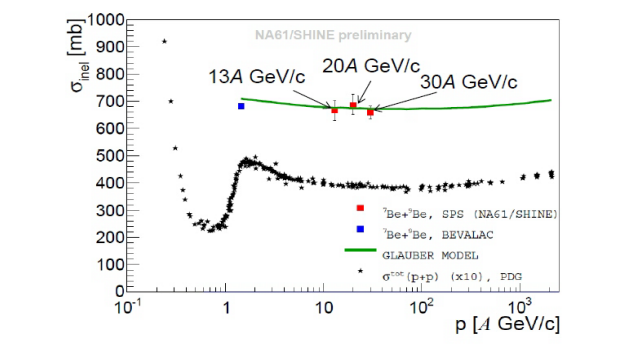}
\caption{\label{Fig.3}
  The inelastic cross section for ${}^7Be + {}^9Be$ collisions.  It is well reproduced by the Glauber model\cite{Broniowski:2009}}
\end{figure}
\subsubsection*{}
Pion spectra for ${}^7Be + {}^9Be$ collisions
 \begin{figure}[!h]
\includegraphics[width=\textwidth]{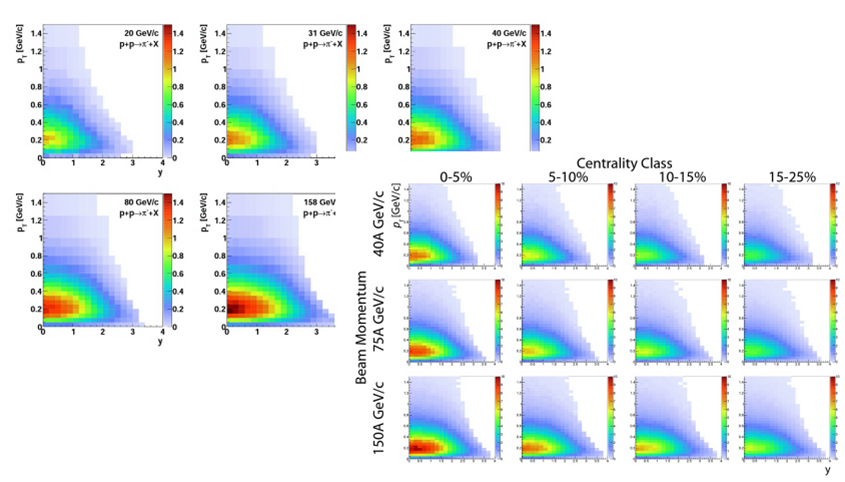}
\caption{\label{Fig.4}
  Precise measurements of $\pi^-$ in $p+p$ and ${}^7Be + {}^9Be$ collisions}
\end{figure}

\subsubsection*{}
\break
\subsubsection*{}

Pion rapidity distribution for ${}^7Be + {}^9Be$ collisions

\begin{figure}[!h]
\includegraphics[width=\textwidth]{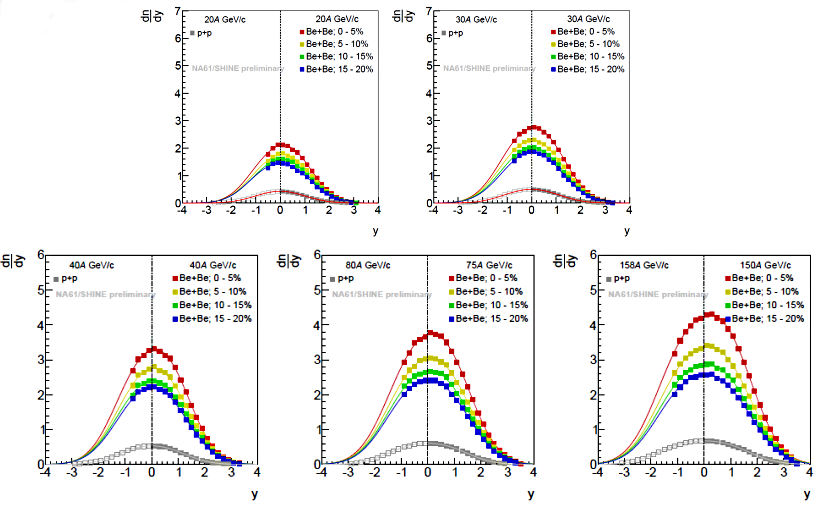}
\caption{\label{Fig.5}
  Pion rapidity distribution of $\pi^-$ in $p+p$ and ${}^7Be + {}^9Be$ collisions}
\end{figure}

\subsubsection*{}

Inverse slope parameter $T$

\begin{figure}[!h]
\includegraphics[width=0.9\textwidth]{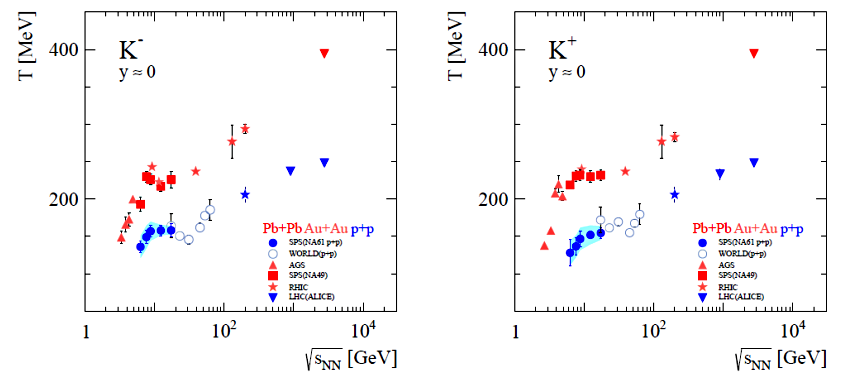}
\caption{\label{Fig.6}
Inverse slope parameter $T$  exhibits rapid changes in
the SPS energy range - also in $p+p$ collision}
\end{figure}

\begin{figure}[!h]
\includegraphics[width=0.5\textwidth]{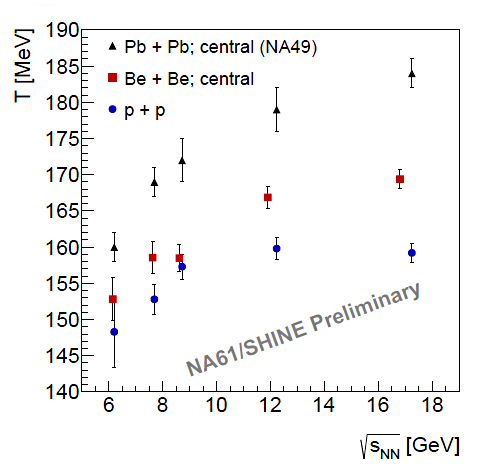}
\caption{\label{Fig.7}
Inverse slope parameter $T$  in ${}^7Be + {}^9Be$ larger than in $p+p$ collision indicates possibility of the transverse collective flow}
\end{figure}

\subsubsection*{}

Transverse momentum fluctuations
\begin{figure}[!h]
\includegraphics[width=0.7\textwidth]{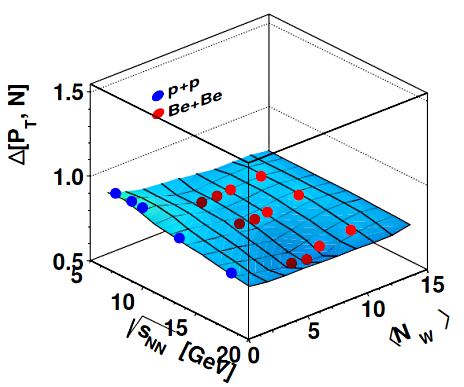}
\caption{\label{Fig.8}
No critical fluctuations neither  in ${}^7Be + {}^9Be$ nor in $p+p$ collision }
\end{figure}

\vfill

\ack

This work was supported in part by the Polish National Science Center (NCN)
under contract No. DEC-2012/04/M/ST2/00816
\\[5mm]

\end{document}